# Service Networks Monitoring for better Quality of Service

Tehreem Masood, Chantal Bonner Cherifi, Néjib Moalla
University of Lyon 2, DISP Laboratory, Lyon, France
Tehreem.Masood@univ-lyon2.fr, chantal.bonnercherifi@univ-lyon2.fr, Nejib.Moalla@univ-lyon2.fr

*Abstract*— **Today, the deployment of Web services in many enterprise applications has gained much attention. Service network inhibits certain common properties as they arise spontaneously and are subject to high fluctuation. The objective of consumer is to compose services for stable business processes in coherence with their legacy system capabilities and with better quality of services. For this purpose we have proposed a dynamic decision model that integrates several performance metrics and attributes to monitor the performance of service oriented systems in order to ensure their sustainability. Based on the available metrics, we have identified performance metrics criteria and classified into categories like time based QoS, size based QoS, combined QoS and estimated attributes. Then we have designed service network monitoring ontology (SNM). Our decision model will take user query and SNM as input, measures the performance capabilities and suggests some new performance configurations like selected service is not available, physical resource is not available and no maintenance will be available for the selected service for composition.**

## INTRODUCTION

Service-Oriented Computing (SOC) increasingly gains motivation in both industry and academia as a means to develop adaptive distributed software applications in a loosely coupled way. The motivation behind SOC is the idea that businesses offer their application functionality as services over the Internet and other users or companies can integrate and compose these business services into their applications [1]. Web services became very important during the past few years. It is a software system designed to support interoperable interaction between different applications and different platforms [2]. Web services use standards such as Hypertext Transfer Protocol (HTTP), Simple Object Access Protocol (SOAP) [3], Universal Description, Discovery, and Integration (UDDI) Web Services Description Language (WSDL) [4] and Extensible Markup Language (XML) for the communication between web services through internet.

Web services flow specifications like business process execution language for Web services (BPEL4WS) [5] and Web services choreography interface (WSCI) have also been discussed in literature [5]. Since business requirements are becoming the major driving force for creating Web services research topics to support the business process integration, collaboration, and management, the business context should be captured and transmitted into appropriate partners [5].
There are two levels of performance problems of Web services, namely system level which is related to SOAP and XML, and server level which is related to the processing of SOAP requests at the server side [6]. If we deal with them properly, it will definitely provide a more efficient and scalable structure in terms of performance for deploying and running Web services. Web services are supposed to be a source of generating increased returns for enterprises by exposing the legacy enterprise applications to a wide range of other applications on different platforms [7]. It is important to address the better quality of service technique that makes efficient use of available resources.

Business service developers are just to assemble a set of appropriate web services to implement the business tasks. Business applications are no longer written manually. For example [8], a client requirement can be expressed as a sequence diagram in UML. It is composed of several sub-functional modules or abstract services. Each abstract service is associated with a web service community which contains several existing web services with the same functionality. The process of selecting a service from a web service community for an abstract service by quality of service attributes is called the local selection. As a task presented by the service composition can be explained by a significant number of combinations.

The objective of consumer is to compose services for the business collaboration. Constraints are not to impact the performance of legacy systems as well as the service composition for stable business and with better quality of services. For this purpose we have proposed a dynamic decision model that integrates several performance metrics and attributes to monitor the performance of service oriented systems in order to ensure their sustainability. Based on the available metrics, we have identified performance metrics criteria and classified into categories. Then we have designed service network monitoring ontology (SNM). Our decision model will take user query and SNM as input, measures the performance capabilities and suggests some new performance configurations like selected service is not available, physical resource is not available and no maintenance will be available for the selected service for composition.

The remaining paper is organized as follows: Section II includes related work in the area of performance of web services and modern technologies of web services. Section III discusses the proposed approach. In the end, conclusion of the paper is presented in Section IV.

## RELATED WORKS

In this section we have classified the related work into three categories. System level performance, server level performance and different web services architectures. These three types of techniques will help reader to understand about the existing work of different performance levels like domain, node, service, server and messaging. We are concerned with all these performance levels.

### 1. System Level Performance

The System Level covers domain level, messaging or service level, node level and service level performance. In this category we have discussed some techniques that are related to system level performance.

Z. Tari et al. [10] proposed a benchmark of different SOAP bindings in wireless environments. Its configuration and results can serve as a standard benchmark for other researchers who are also interested in the performance of SOAP bindings in wireless networks. Three sets of experiments were carried out: loopback mode, wireless network mode and mobile device mode. The experimental results show that HTTP binding inherits very high protocol overhead (30%–50% higher than UDP binding) from TCP due to the slow connection establishments and tear-down processes and the packet acknowledgement mechanism. UDP binding has the lower overhead because it does not require establishing connections before transmitting datagrams and does not address reliability. This results in a reduction in the response time and an increase in the total throughput.

Z. Tari et al. [11] proposed a similarity-based SOAP multicast protocol (SMP) which reduces the network load by reducing the total generated traffic size. It is based on the syntactic similarity of SOAP messages. In particular, the SMP reuses common templates and payload values among the SOAP messages and only sends one copy of the common part to multiple clients. SMP makes use of the commonly available WSDL description of a SOAP Web service when determining the similarity of response messages. For messages that are highly similar, instead of generating messages with duplicated similar parts for different clients, the duplicated parts are reused for multiple clients and are sent only once from the source .

Z. Tari et al. [12] proposed a tc-SMP1 technique which is an extension of SMP, which is the traffic-constrained similarity-based SOAP multicast protocol (tc- SMP) is proposed here. Two algorithms, greedy and incremental approaches, are described to address this problem. Both tc-SMP algorithms aim at minimizing the total network traffic of the whole routing tree every time a new client is added to the tree. Two heuristic methods are also proposed for these algorithms to assist in choosing the order of clients being added to the tree. In general, the performance improvement of tc-SMP is about 30% higher network traffic reduction than SMP at a small expense of up to 10% rise in the response time.

Comparison of the system level performance techniques is described in Table I. Parameters used in the comparison table are response time, throughput, network traffic, binding type and similarity matching.

**Response time:**
It is also called latency. It is the time perceived by a client to obtain a reply for a request for a web service. It includes the transmission delays on the communication link. It is measured in time units

**Throughput:**
The number of requests executed per unit of time. For web service users it can be measured by requests per seconds or number of operations per second

**Network Traffic:**
The total network traffic for communication scheme or session which is the number of bytes transferred during the communication

**Binding Type:**
Binding type is the type of protocol used for binding. Either it is http, udp or more

**Similarity Matching:**
Similarity matching is the type of matching like syntactic or semantic. NA in Table 1 means that this parameter is

TABLE I
COMPARISON OF SYSTEM LEVEL PERFORMANCE TECHNIQUES

| Techniques | | Parameters | | | | |
|---|---|---|---|---|---|---|
| | | Response Time | Throughput | Network Traffic | Similarity Matching | Binding Type |
| SOAP Binding | SOAP over HTTP | Large | Small | Large | HTTP | NA |
| | SOAP over TCP | Medium | Medium | Large | TCP | NA |
| | SOAP over UDP | Small | Large | Small | UDP | NA |
| The Use of Similarity & Multicast Protocols to Improve Performance SMP [7] | | Medium | Medium | Medium | SOAP Over HTTP | Syntactic |
| Network Traffic Optimization tc-SMP1 [8] | | Large | Large | Small | SOAP Over HTTP | Syntactic |

not applicable to the corresponding technique

### 2. Server Level Performance

In this category we have discussed some techniques that are available for server level performance.

A series of advanced task assignment policies: TAGS (Task Assignment by Guessing Size), TAGS-WC (TAGS with Work Conserving), TAPTF (Task Assignment based on Prioritizing Traffic Flows), TAPTF-WC (TAPTF with Work Conserving), and MTTMELL (Multi-Tier Task Assignment with Minimum Excess Load) Policy.
Multi-level Time Sharing (MLTP) Policy investigated time sharing under more general setting where amount of time service time (quantum) allocated on levels using a random variable.

### 3. Different Web Services Architectures

In this category, we have discussed various web services architectures that cover different level of performance like messaging or service level and node level performance.

Zhou et al. [9] proposed UX which is an extended UDDI. It assesses previous and current service usage for the future service selection. With the analysis of the network model, the condition of service requester's connection is recorded by the server to enable better predictions in a future service's request. Instead of the QoS description published by service provider, QoS feedbacks made by service requesters are used to generate summaries for the invoked services. These summaries are then used to predict the services' future performance. A general federated service is designed so that server nodes can be administratively federated across network boundaries. Based on this federated service, lookup interface is provided on a UX server that facilitates the discovery between different registries and the exchange of service QoS summaries.

Bertino et al. [13] proposed an approach based on Merkle hash trees, which provides a flexible authentication mechanism for UDDI registries. They have claimed two relevant benefits. The first is the possibility for the service provider to ensure the authenticity and integrity of the whole data structures by signing a unique small amount of data, with the obvious improvement of the performance. The second benefit regards browse pattern inquiries is that they return the overview information taken from one or more data structures. According to the UDDI specification, in such a case if a client wishes to verify the authenticity and integrity of the answer, it must request the whole data structures from which the information are taken.

Curran and Gallagher [14] proposed a framework called Webber, which provides the services necessary for supporting new communication protocols and qualities of service. Webber consists of a set of Java classes for representing the uniform resource locators, protocol stacks, the framework API and SOAP. The abstraction is analogous to that of the various broadcast media in everyday use, such as newspaper, radio and TV corresponding to text, audio and video components contained in multimedia applications. Webber fragments the various media elements of a multimedia application and 'broadcasts' them over separate channels to be subscribed to at the receiver's own choice.

Mateos et al. [15] proposed a scheme named as MoviLog. MoviLog is a platform for building the intelligent mobile agents, based on a strong mobility model, where agents' execution state is transferred during migration. MoviLog is an extension of JavaLog a framework for an agent-oriented programming. In order to provide mobility across sites, each MoviLog host has to execute a MARlet which is a mobile agent resource. A MARlet is a Java servlet that encapsulates a Prolog inference engine and provides services to access it. In this way, a MARlet represents an execution environment for mobile agents, or brainlets in MoviLog terminology. In this sense, a MARlet can be used as an inference server for agents and external Web applications. This mechanism states that when certain predicates previously declared in the code of a Brainlet fail, MoviLog transparently moves the Brainlet and its execution state to another site that contains definitions for that predicate, thus making local use of those definitions later.

Comparison of some of web services architectures level technologies of web services discussed in this paper is described in Table II. Parameters used in the comparison table are response time, reliability, standard used and cost. Response time has already been explained. Other parameters are explained below.

**Reliability:**

Reliability corresponds to the likelihood that the service will perform when the user demands it and it is a function of the failure rate. Each service has two distinct terminating states: One indicates that a Web service has failed or aborted, the other indicates that it is successful or committed

**Standard Used:**

It describes the type of standard used in the corresponding technique

**Cost:**

Cost represents the cost associated with the execution of the service.

TABLE II
COMPARISON OF SOME WEB SERVICE ARCHITECTURES

| Techniques | Parameters | | | |
|---|---|---|---|---|
| | Reliability | Response Time | Standard | Cost |
| QoS-Aware Web Services(UX) [5] | Medium | Large | UDDI | High |
| Merkele [9] | High | Large | UDDI | High |
| Webber [10] | High | Large | SOAP | High |
| MoviLog [11] | High | Small | OWL S | High |

## III. RESEARCH CHALLENGES

There are several challenges that have been addressed in the literature in order to monitor the performance of web service network at different levels. All the levels are shown in Figure 1.

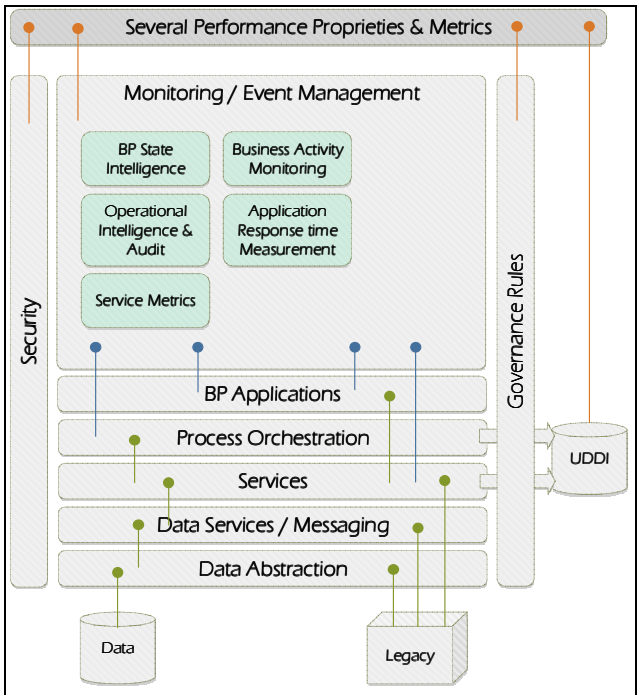

Figure 1. Research Challenges of Performance of SOA

Data abstraction layer is used to query data from the database and the historical information about data and components which is available in the legacy. Data abstraction layer will provide the data services to the messaging layer. Once data services are gathered then messaging layer provides the ability to perform the necessary message transformation to connect the service requestor to the service provider and to publish and subscribe messages and events asynchronously. In this way services are published in the service layer. Then we have the whole service oriented architecture in the process orchestration layer. All this information is stored in the UDDI. The next layer of BP application provides the service composition. Governance rules are the set of policies like service will be available for one year etc. Security is used to provide some integrity to the system like authentication with the help of user name and password. Service metrics are the parameters in order to guarantee the performance of web services. For example in the information technology infrastructure library (ITIL) there are 5 sub-categories and more than 100 metrics available for service support process and 5 sub-categories and more than 50 metrics available for service delivery processes. Operational intelligence and audit component is used to add some parameters in the query to measure the performance of the services. BP state intelligence is used to provide some intelligence or flags to measure the performance of BPEL. BAM (Business activity monitoring) and ARM (Application response time measurement) are the infrastructures that will help to measure different quality metrics.

This shows that different kinds of technologies and infrastructures are available that can be utilized to measure the performance of service network at different levels. But they are not effectively utilized in order to provide better quality of service in service oriented architectures (SOA). We can measure generic traces for web service network monitoring like daily, weekly and monthly trend in the value fluctuation, loss or error of service, un availability of service, SLA violations, resources are not available and new service requirement.

## IV. PROPOSED APPROACH

First of all, the identification of performance criteria from the available performance metrics has been performed. Several performance metrics are available for decision. In Figure 2, we have shown different levels of performance with the help of lending example. Customer wants to take loan from the loan organization. Loan organization is the domain that has many services like receive application, check credit, negotiate loan, close loan and book loan. At the node level we have shown the composition of services like customer accounting is the composition of check credit and book loan. Similarly credit administration is the composition of check credit and negotiate loan. Service messaging is related to the protocols used. Loan organization system is the server that manages the physical resources as well as receives and modifies application. We have shown all these performance levels in our ontology discussed in Section IV.

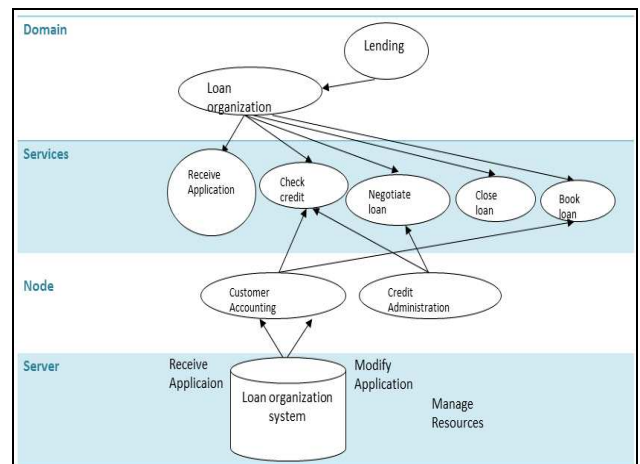

Figure 2. Example

The next step is the classification of these metrics and attributes based on some criteria. Qos metric values are gathered from the metrics database and used as attributes of a record of the Matching set. Then, the choice of KPI for the evaluation based on the user requirement by selecting KPI target value and analysis period in the structuring phase. The Predicate of KPI metric is classified as valid or violated. Then a decision model have been proposed that will generate answers like the selected service is not available for an additional consumption, no physical resources are available to support the new deployment, no maintenance will be available for the selected service for composition and security compliance problem in the deployment. All these steps are shown in Figure 3.

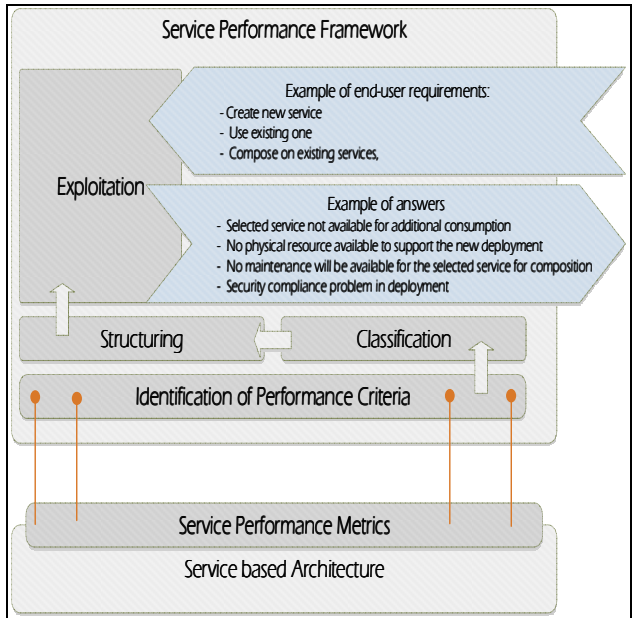

Figure 3. Proposed Approach

### 1. Classification of Metrics

Following is the classification of metrics from the available metrics of the information technology infrastructure library (ITIL)

A. *Time based*
Availability: Total down time per service
Delay - Downtime divided by Uptime
Response times per incident
Actual availability compared with SLA requirements

B. *Size based*
Reliability – loss or error – Number of successful invocations divided by total

C. *Combined (both time and size based)*
Bandwidth – Tasks per time unit and average data blocks per time unit
Throughput – number of operations per second

D. *Estimated attributes (historical or prediction)*
CPU load, network load, free RAM, Free Disk space etc

### II. Structure of Metrics

Structure of metrics and key performance indicators has been designed with the help of ontology.

A. *Service Network Monitoring (SNM) Ontology:*
Service domain concept, QoS (quality of service), performance levels and KPI have is a relationship with the system. Then we have defined that service, service provider, service consumer and service host have is a relationship with service domain concept. Likewise time based QoS, size based QoS, combined QoS and estimated attributes have is a relationship with QoS as explained step I. Domain, node, service, service messaging and server have is a relationship with performance levels that have already been explained in section III. Similarly response time, delay, error, loss, SLA, number of operations per second and average data blocks per time unit have is a relationship with KPI(key performance indicators) as explained in the beginning of classification of metrics.

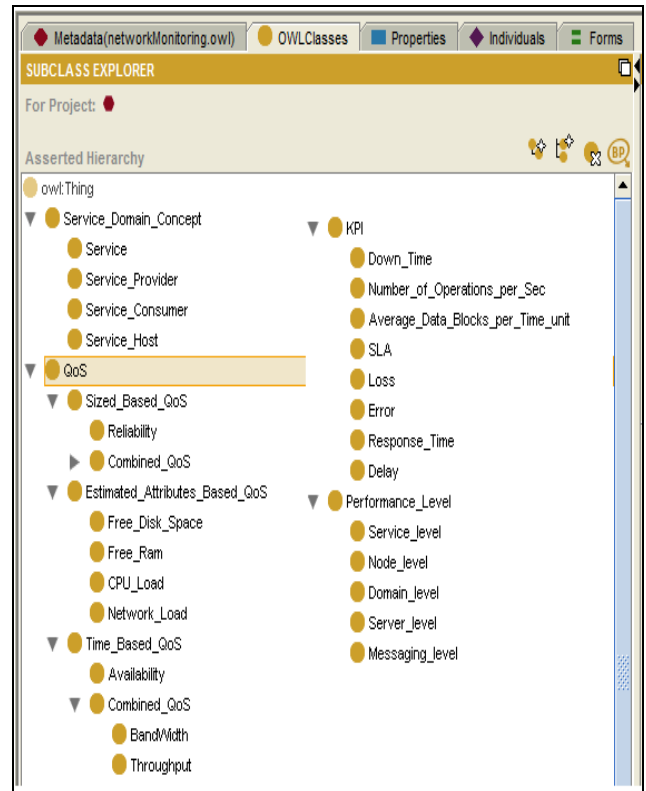

Figure 4. Service Network Monitoring (SNM) Ontology

Different key performance indicators of the metrics from the classification step based on the example discussed in section III are shown in Figure 5.

| Domain | Node | Server | Service | Messaging |
|---|---|---|---|---|
| Loan organization | Customer accounting | Loan Organization Server | Check credit | Service messaging |
| Response Time | Loss | Network load | Response time | Response time per incident |
| Total Number of Users | Error | CPU load | Delay | Binding type |
| 5 services | Delay | Requests Completed | Actual availability compared to SLA | No of bytes transferred |
| 2 nodes | Response time | Average Request | No of operations per sec | No of operations per sec |
| Delay | | Processing Time (seconds) | Loss | Average data blocks per time unit |
| | | Available memory | Error | Loss / error |

Figure 5. Key Performance Indicators

### III. Decision Model

Decision model is explained with the help of pseudocode as shown in Figure 6. It takes ontology and user query as inputs and outputs will be selected service is not available, physical resource is not available for new deployment, maintenance of selected service is not available for composition and security compliance problem. First of all service has been selected that matches to the user query.

If service is available then a function named as check KPI is called. If service is not available then a function named as check services for composition has been called. In this function all the services have been checked for service composition based on user query. If service composition is possible then a function named as check KPI is called. If service composition is not possible then a function named as create new service is called. In check KPI function all the KPI have been checked as shown in Figure 4. For each KPI, a comparison has been performed with SLA (service level agreements). If it is matched with the service level agreements then use that service as it is. If it is not matched then a new function named as check service status has been called. In check service status function service status has been checked. If service status is success then three functions have been called in order to reach to the cause of the problem. The three functions are check protocol, check server and check nodes. In check protocol function details of the protocol that have been used to bind the service needs to be checked. If it is the source of problem then we need to change the messaging protocol. In check server, all the available resources that are provided at the server level have been checked. If they are the source of problem then we need to change the resources as shown in the estimated attribute node of the ontology. In check node function, service composition has been checked. If it is the source of problem then we need to change the service composition or create a new service.

```
Begin
       Input: User query, Ontology
       Output:
       Selected service ≠ Available
       Physical resource ≠ Available
       Maintenance of selected service ≠ Available for
       composition
       Security compliance problem
       Step 1:
       Select service that matches user query.
       Step 2:
       If Service = Available
       Then
                Check KPI
       Else
                Check for services for composition
       Step 3:
       Check for services for composition
       For each service = Available for composition
       Then
                Check KPI
       Else
                Create new service
       Step 3:
       Check KPI
       Step 4:
       For each KPI
                Compare from SLA
       If it is matched
       Then
                Use service or service composition as it is
       Else
                Check service status
       Step 5:
       Check service status
       If service status is success
       Then
                Check protocol
                Check server
                Check nodes
       Else
                Repeat step 2 for checking availability of other
                services
End
```

Figure 6: Pseudocode of Proposed Approach

their sustainability. First of all we have classified the performance metrics from the available metrics of the information technology infrastructure library (ITIL). Then we have designed our system ontology to show the relationships of all the concepts that we have used in our decision model. Finally a decision model has been designed in the form of pseudocode. Our next step will be to validate first case using "Oracle® Content Services Administrator"

7[20] Systems Management: Application Response Measurement (ARM) API. Technical Standard C807, The Open Group, July 1998

[21] P. Bhoj, S. Singhal and S. Chutani. "SLA Management in Federated Environments" In proceedings of the sixth IFIP/IEEE International Symposium on Integrated Network Management, Boston, MA, 1999
7